\documentclass[twoside]{ilcws08}
\usepackage[latin1]{inputenc}
\usepackage[dvips]{graphicx,epsfig,color}
\usepackage{wrapfig,rotating}
\usepackage{amssymb,amsmath,array}

\usepackage{epsf,url}
\usepackage{epstopdf}

\newcommand{\dream} {{\sc dream~}}
\newcommand{\C}     {\v{C}erenkov }

\begin{document}
\title{Particle Identification in 4th}
\author{John Hauptman
\thanks{This work was supported by the DoE Grant at ISU, by
the ADR program of DoE at TTU, the LCRD program at U Oregon, and INFN, Lecce.}
\vspace{.3cm}\\
Iowa State University,  Department of Physics and Astronomy \\
Ames, IA 50011  USA, and
\vspace{.1cm}\\
Texas Tech University,  Department of Physics \\
Lubbock, TX  79409  USA\\
}

\maketitle

\begin{abstract}
We describe 12 measurements in the 4th detector that yield particle identification information.  Seven of these have been demonstrated with test beam data from the {\sc dream} collaboration,  one demonstrated in cosmic muon test data, one verified in ILCroot, and the remaining three will be tested in ILCroot.  Not all are independent, but as a whole they cover all partons of the standard model.
\end{abstract}

\noindent{\bf The importance of particle identification}

Physics measurements often depend on the efficiency and the purity of an event ensemble, and the efficiency is a product of several small efficiencies for the isolation and 
identification of each event feature.   The 4th detector has been designed from the beginning with particle identification in mind, knowing that we may be seeking small signals and that,
in this sense, identification efficiency is equivalent to luminosity.

The 4th detector is unique in several respects and is rich in particle identifications measurements some of which are new in high energy physics.    The dual-readout calorimeters are responsible for the majority of the ``electromagnetic'', ```muonic'', or ``hadronic'' tags;  the cluster-counting CluCou chamber provides $e-\mu-\pi^{\pm}-K-p$ discrimination in the few-GeV region by a precision measurement of specific ionization; the resolutions of both the tracking chamber and calorimeters through ILCroot are responsible for tagging $W \rightarrow jj$ and $Z \rightarrow jj$ decays and for $e-\gamma$ discrimination; and, the time resolutions of the optical calorimeters (both fiber and crystal) yield the time-of-flight resolutions.

\begin{wrapfigure} {l} {0.65\columnwidth}
\centerline{\includegraphics[width=.6\columnwidth]{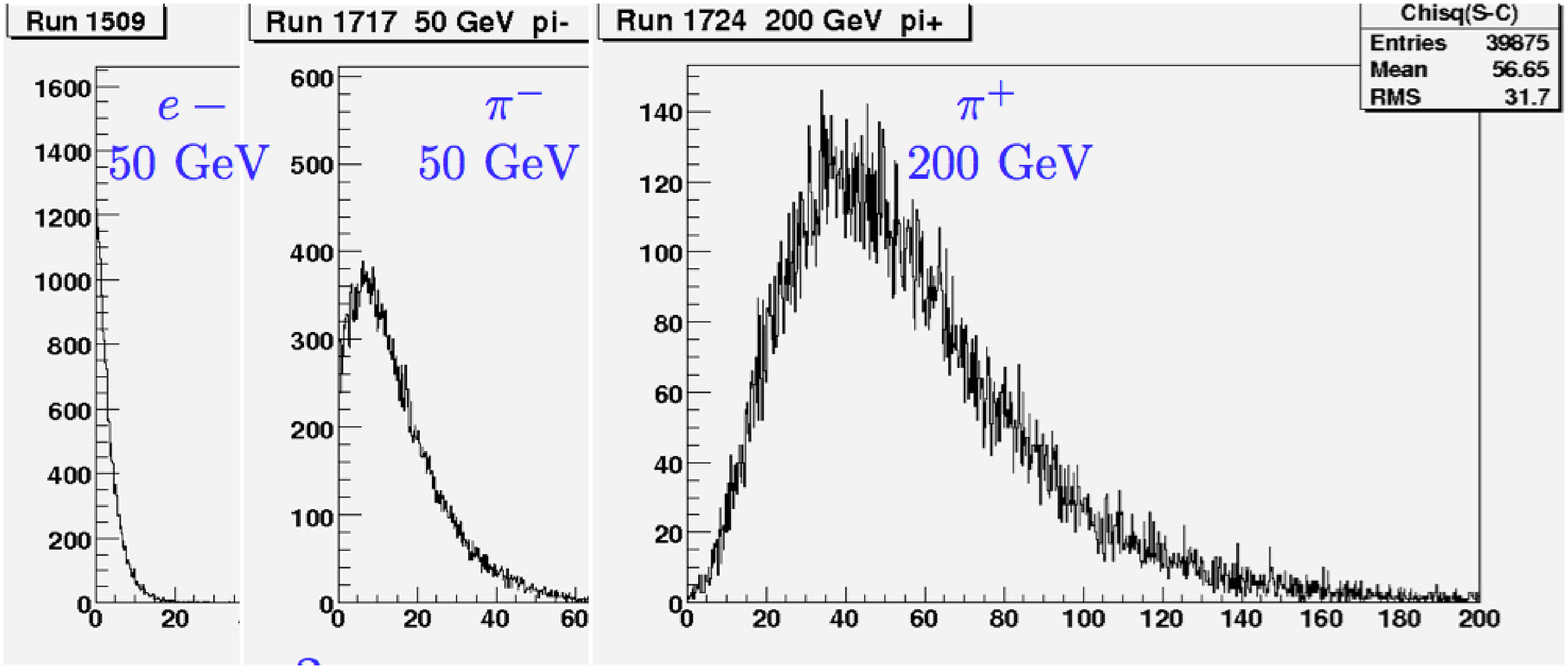}}
\caption{
}\label{Fig:Chisq}
\end{wrapfigure}

\medskip \noindent{\bf Electromagnetic {\it vs.} hadr\-on\-ic}

The leptonic decays of $W$ and $Z$ and, expectantly, the decays of more exotic particles, yield isolated $e^{\pm}$, $\pi^{\pm}$ and $\mu^{\pm}$ from such decays as $W \rightarrow e \nu$, $Z \rightarrow \tau^+\tau^-$, and $\tau \rightarrow \pi \nu, e^{\pm} \nu \nu, \mu^{\pm} \nu \nu$, in addition to leading particles from jets.  It is essential that these species be cleanly separated from each other in a detector.

\medskip \noindent {\bf  (i) S {\it vs.} C:}
In the dual-readout calorimeters, both the scintillation signal ($S$) that measures all charged particles including nuclear spallation protons, and the \C signal ($C$) that measures predominantly the $e^{\pm}$ generated in electromagnetic showers from $\pi^0 \rightarrow \gamma \gamma$ decay, are measured simultaneously.   Since $S$ and $C$ respond differently to electromagnetic showers and hadronic showers a plot of $S ~vs.~ C$ yields a clear two-dimensional separation of electromagnetic and hadronic showers.   Specifically, for electromagnetic showers, $S \approx C$, whereas $S \gtrsim C$ for hadronic showers.

\medskip \noindent {\bf (ii) Fluctuations in $(S-C)$:}
The $S$ and $C$ signals in {\bf (i)} are the sums of the channels that constitute the shower.  A nearly independent discriminator is the channel-by-channel deviations of $S$ from $C$, for which the chi-squared
$\chi^2_{S-C} = \sum_{k} \left[ (S_k - C_k)/\sigma_k \right] ^2$
is small for electromagnetic showers and large for hadronic showers, with
$\sigma_k$ the expected rms variation of $(S-C)$, and $k$ the channel index.  This chi-squared is shown in Fig. \ref{Fig:Chisq} for electrons and pions at 50 GeV and pions at 200 GeV.\cite{dream-h}

\begin{wrapfigure}{l}{0.65\columnwidth}
\centerline{\includegraphics[width=0.6\columnwidth]{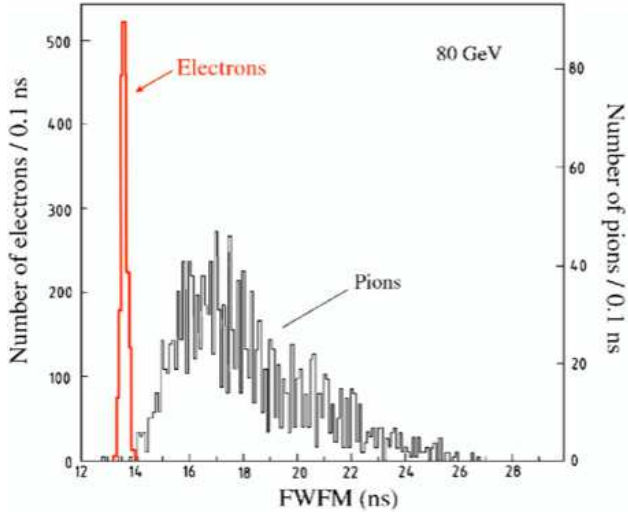}}
\caption{PMT pulse duration for $e$ and $\pi^{\pm}$ at 80 GeV.}\label{Fig:SPACAL}
\end{wrapfigure}

\medskip \noindent {\bf  (iii) Time-history of light production:}
A third statistic, independent of {\bf (i-ii)}, is contained in the time structure of the scintillation signal, $S(t)$, digitized in approximately 1ns bins.  The space-time structure of an electromagnetic shower is essentially a velocity-$c$ pancake of particles that passes through the calorimeter medium leaving behind optical photons that travel at $v \approx c/n$.  Since electromagnetic showers are all quite similar in space-time, their photon arrival time distributions are similar.  The duration of the scintillation pulse (the width of the pulse at one-fifth maximum) is shown in Fig. \ref{Fig:SPACAL} for electrons and pions at 80 GeV.\cite{SPACAL}

\medskip \noindent{\bf   Muonic {\it vs.} non-muonic}

\begin{wrapfigure}{r}{0.5\columnwidth} 
\centerline{\includegraphics[width=0.45\columnwidth]{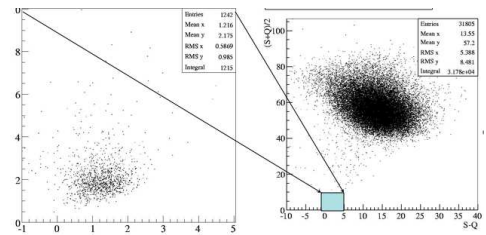}} 
\caption{(S+C) {\it v} (S-C), 80 GeV $\mu^{\pm}$ and $\pi^{\pm}$.}
\label{Fig:mu}
\end{wrapfigure}
\noindent {\bf (iv)} The dual-readout fiber cal\-or\-i\-meter provides a positive identification of a $\mu^{\pm}$ shown in Fig. \ref{Fig:mu} as 
particles with $(S-C) \approx 1$ GeV {\it independent of the degree of radiation} inside the calorimeter, and $(S+C)/2 \approx$  energy radiated inside calorimeter.

\begin{figure}[hbtp]

  \centerline{\hbox{ \hspace{0.0in} 
    \epsfxsize=2.5in
    \epsffile{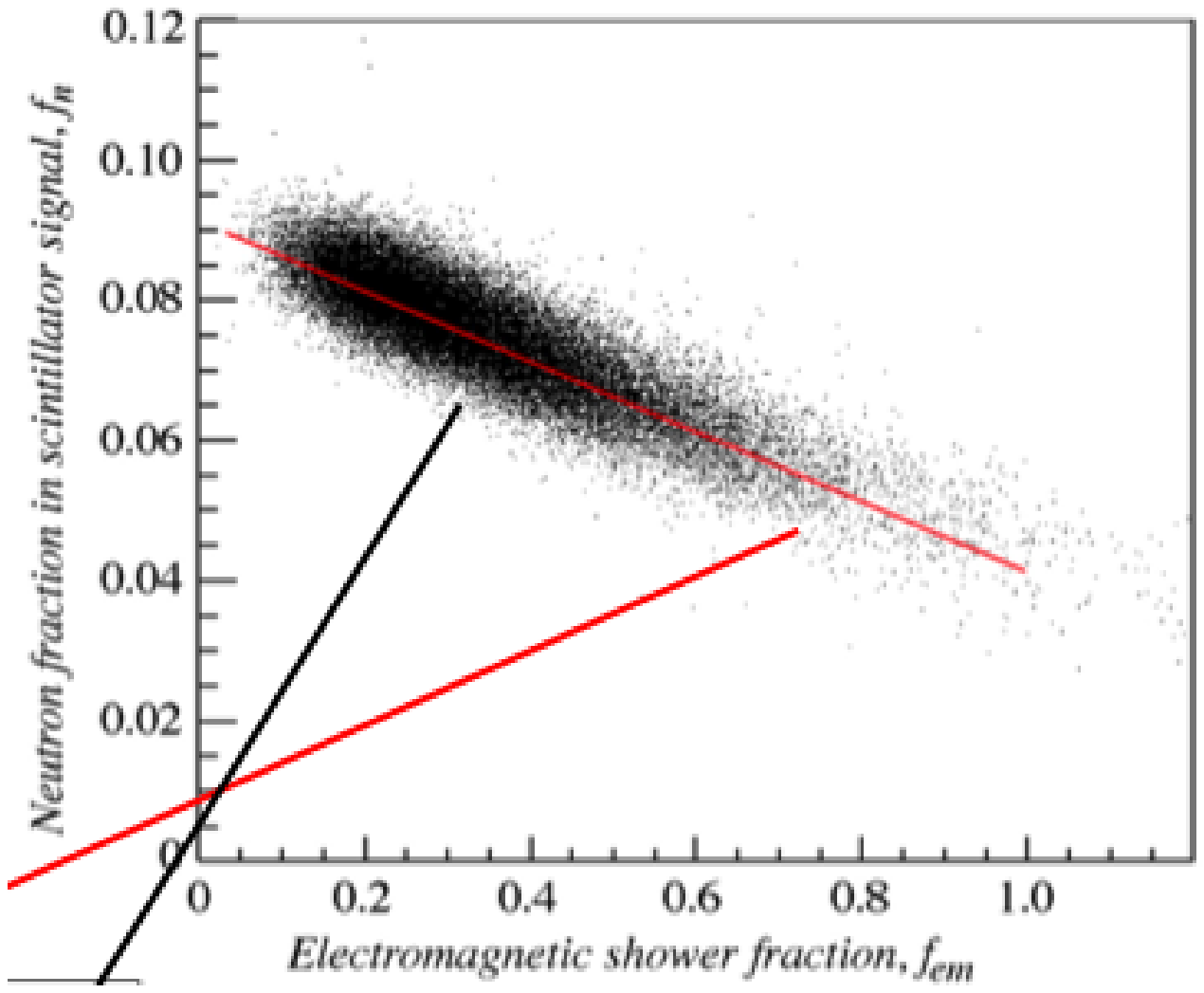}
    \hspace{0.25in}
    \epsfxsize=2.0in
    \epsffile{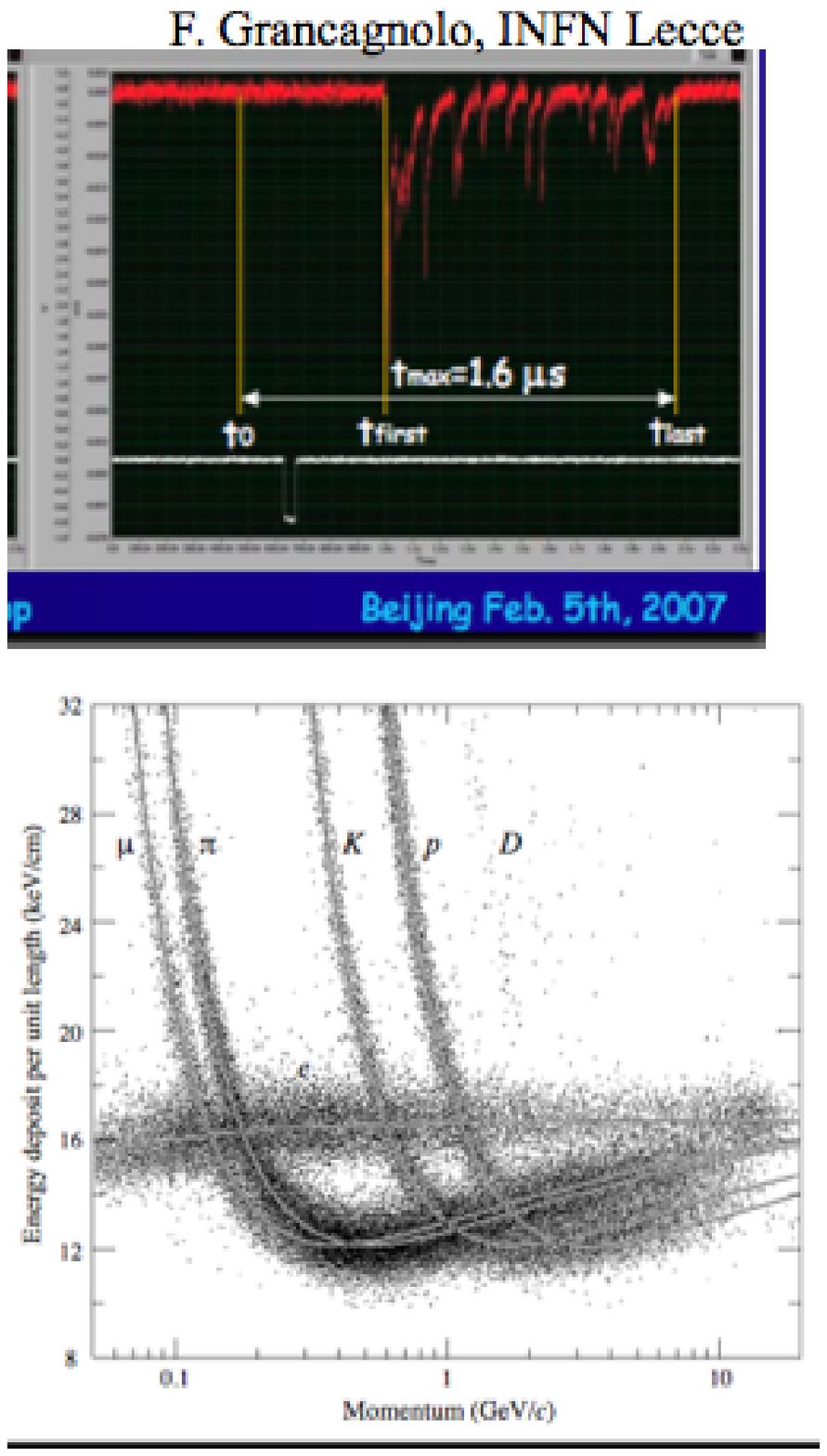}
    }
  }

  \hbox{\hspace{1.35in} (a) \hspace{2.10in} (b)} 
  \vspace{9pt}

  \centerline{\hbox{ \hspace{0.50in}
    \epsfxsize=2.5in
    \epsffile{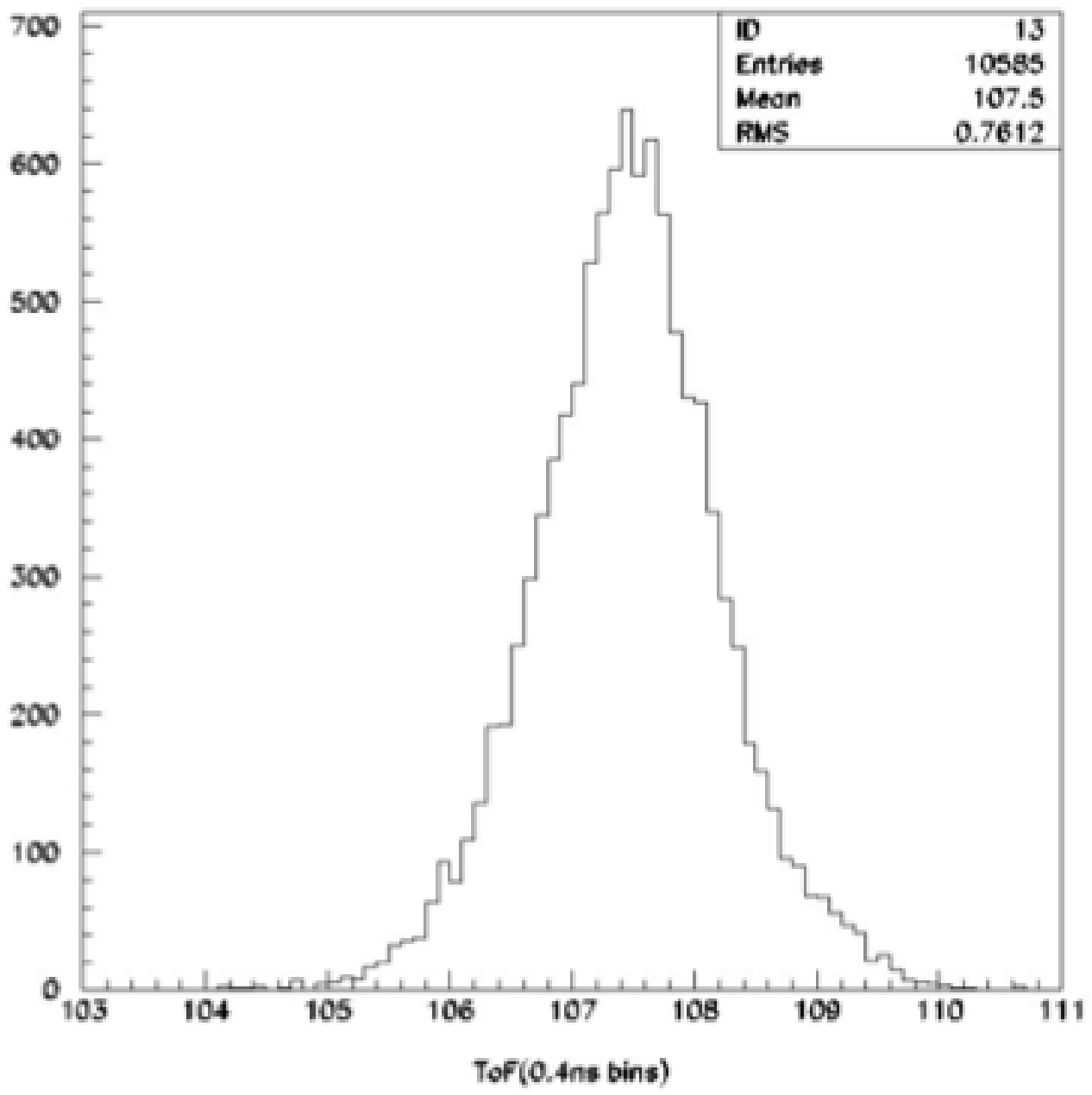}
    \hspace{0.25in}
    \epsfxsize=2.9in
    \epsffile{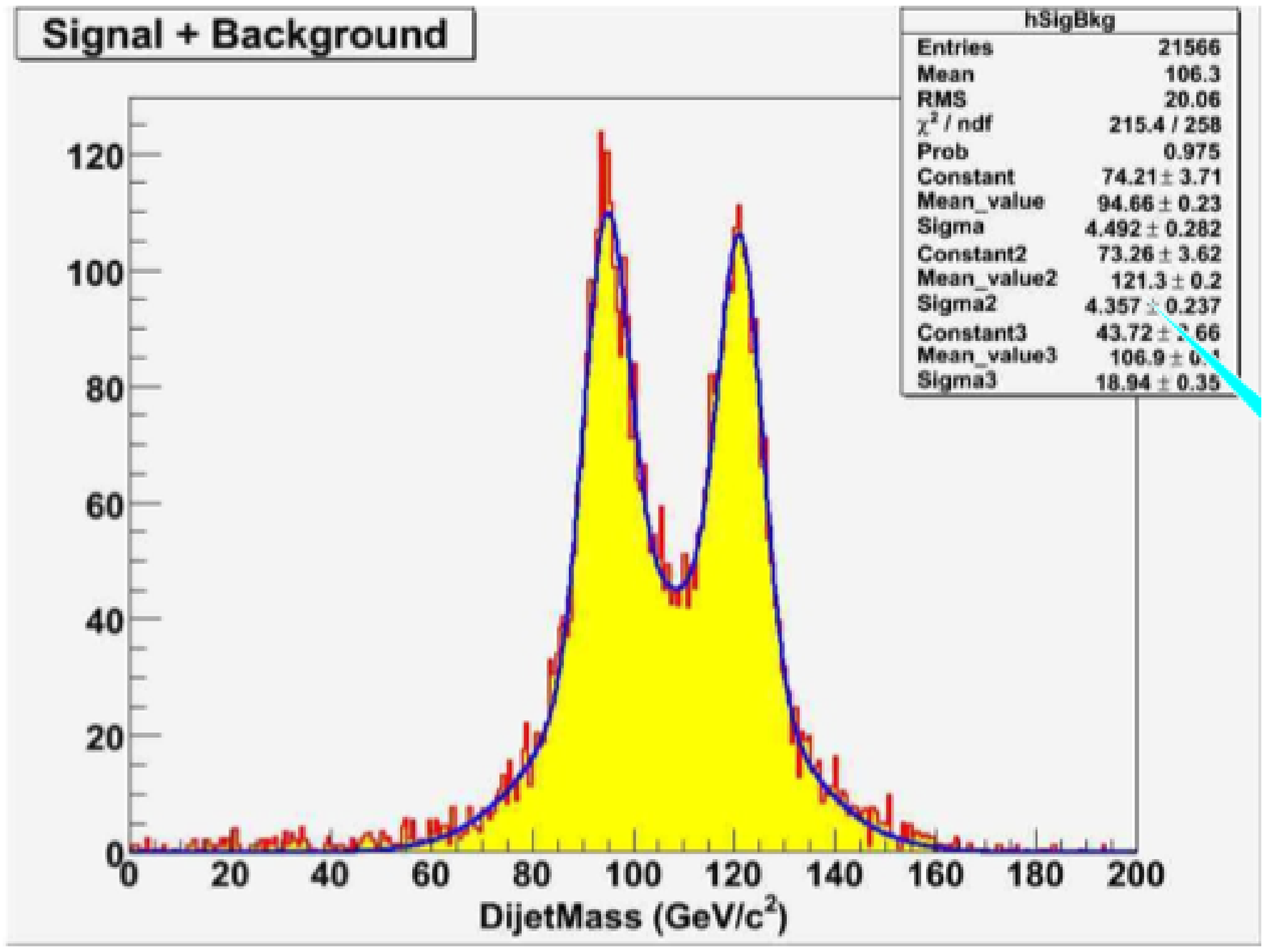}
    }
  }

 \hbox{\hspace{1.5in} (c)       \hspace{2.10in} (d)} 

  \caption{(a) The neutron fraction, $f_n$, anti-correlated with the electromagnetic fraction, $f_{EM}$, in \dream data.   (b) Upper panel is an oscilloscope trace of electron clusters in CluCou test; lower panel is $dE/dx$ measurement from PEP4 TPC.   (c) Time-of-flight resolution in Cerenkov light from \dream;  (d) the two-jet mass distribution from $e^+e^- \rightarrow HZ \rightarrow c\bar{c} \nu \nu$ and the background process $e^+e^- \rightarrow ZZ \rightarrow q\bar{q} \nu \nu$. }
  \label{Fig:2x2}
\end{figure}

\medskip \noindent{\bf  Hadronic {\it vs.} non-hadronic}

\noindent {\bf  (vi)} The time-history of the scintillating fibers allows a measurement of the slow MeV neutrons liberated in nuclear break-up that are correlated with the binding energy losses suffered by the hadronic particles of the shower.  These neutrons lose energy in elastic scatters from protons in the scintillating fibers,  are delayed by many tens of nanoseconds, and fill a larger volume than the charged particles of the shower.  The measured neutron fraction, $f_n$, is shown in Fig. \ref{Fig:2x2}(a) against the measured electromagnetic fraction, both measured in the \dream module.\cite{dream-n}

\medskip \noindent{\bf Mass identification}

\medskip \noindent {\bf  (vii) Separation of massive particles from $v\approx c$ particles:} Some theoretical speculations suggest that massive SUSY or technicolor particles with long lifetimes may be produced.  These objects would move into the tracking volume and (likely) 
decay to light-mass particles ($\tau, \mu, e$).  For masses in the region above 100 GeV/c$^2$,
this delayed decay time can be measured by time-of-flight using the \C light in the dual readout calorimeters with a measured time resolution of $\sigma_t \approx 0.30$ ns, shown in Fig. \ref{Fig:2x2}(c).\cite{dream-h}

\medskip \noindent{\bf  (viii) Mass separation by specific ionization:}  
The $t \rightarrow b \rightarrow c \rightarrow s$ decay chain yields $K^{\pm}$,  $\pi, \mu$ and $e$ in the few-GeV region.  These particles, and the decay chain, can be reconstructed by identifying the quark content of each particle by measuring its specific ionization.   We count the ionization clusters on each track {\it without the Landau ionization tail} and  use all clusters on all wires (that is, no truncated mean) and achieve an equivalent $dE/dx$ resolution of 3.5\%.   Measured clusters are shown in Fig. \ref{Fig:2x2}(b).\cite{beijing}

\medskip \noindent {\bf  (ix) $W-Z$ separation by direct two-jet mass resolution:}  
The jet energy resolution achieved in the dual-readout calorimeters\cite{Vito} is $\sigma_E/E \approx 29\%/\sqrt{E} \oplus 1.2\%$ and results in a 2-$\sigma$ separation of $W$ from $Z$ in their hadronic decay final states\cite{Anna} shown in Fig. \ref{Fig:2x2}(d).

\medskip
Particle identifications not yet demonstrated in 4th are {\bf  (x)} $e-\gamma$ separation,  {\bf  (xi)   $\tau \rightarrow \rho \nu \rightarrow \pi^{\pm} \pi^0 \nu \rightarrow \pi^{\pm} \gamma \gamma \nu$} reconstruction, 
and {\bf  (xii)}   $b, c$ quark and $\tau$  lepton impact parameter tagging.
Items {\bf (x-xii)} will be tested in ILCroot.

\medskip \noindent {\bf Acknowledgments}

These results are only possible due to the successful beam tests by the \dream collaboration and the successful simulation and analysis of the whole 4th detector by the Lecce group of C. Gatto, V. Di Benedetto, and A. Mazzacane
(talks given in these proceedings).

\begin{footnotesize}

\end{footnotesize}

\end{document}